\documentclass[runningheads]{llncs}
\usepackage[T1]{fontenc}
\usepackage{graphicx}
\usepackage{booktabs}
\usepackage{amsmath,amssymb}
\usepackage{tabularx}
\usepackage{colortbl}
\usepackage[table]{xcolor}
\usepackage{hyperref}
\usepackage{bbding}

\begin{document}
\title{MVFM-3DAD: Multi-view Flow Matching for 3D Anomaly Detection via Density Proxy Estimation}

\titlerunning{Multi-view Flow Matching for 3D Anomaly Detection via Density Proxy Estimation}
%
\author{
Liangwei Li\inst{1} \and
Lin Liu\inst{1} \and
Jing Zhang\inst{1} \and
Xiaohui Du\inst{1} \and
Ruqian Hao\inst{1} \and
Xinwei Li\inst{2} \and
Hanzhe Liang$^{\dag,}$\inst{2,3}\and
Juanxiu Liu\inst{1}
}
%

\institute{University of Electronic Science and Technology of China, Chengdu 610031, China \and
Shenzhen Audencia Financial Technology Institute,\\ Shenzhen University, Shenzhen, 518060, China \\ \and
Audencia Nantes {\'E}cole de Management, Nantes 44300, Loire-Atlantique, France\\
\email{\{Hanzhe.Liang\}@mbzuai.ac.ae}
}
\maketitle              
\begin{abstract}

In 3D anomaly detection (3DAD), most existing methods rely on Memory bank retrieval or reconstruction. However, memory-based methods are constrained by the coverage of stored normal features, while reconstruction-based methods may learn identity shortcuts that also reconstruct anomalous inputs well. These limitations motivate a density-oriented approach that evaluates whether a test sample follows the learned normal distribution. To this end, we propose \textbf{MVFM-3DAD}, a flow-based framework that reframes 3DAD as density proxy estimation over the normal data distribution. MVFM-3DAD introduces a Bidirectional Geometric Projector (BGP), whose forward process converts irregular point clouds into structured multi-view representations. The Flow-guided Density Proxy Estimator (FDPE) estimates a reference density for each view feature, after which the backward process of BGP maps these multi-view density estimates to their corresponding 3D points. Building on it, anomalous features can be identified by their terminal normality. Unlike conventional flow-based likelihood estimation, our formulation requires neither input reconstruction nor explicit Jacobian evaluation, yielding a simple and efficient anomaly-scoring mechanism. Extensive experiments show that MVFM-3DAD outperforms the strongest competing methods on Real3D-AD and MVTec3D-AD. 

The code is available at \href{https://github.com/lil-wayne-0319/MV3D-AD/}{https://github.com/lil-wayne-0319/MV3D-AD/}.

\end{abstract}

\section{Introduction}\label{intro}

Unsupervised 3D anomaly detection identifies defective objects and localizes anomalous points using only normal training data, making it well suited to industrial inspection, where defects are scarce and diverse~\cite{bergmann2021mvtec}. Existing methods primarily model normality through feature retrieval or reconstruction. Retrieval-based methods store geometric or pretrained descriptors and compare test features against normal prototypes~\cite{horwitz2023back,wang2025m3dm}. Reconstruction-based methods learn a normal shape prior by recovering geometry and use reconstruction residuals as anomaly evidence~\cite{li2024towards,zhou2024r3d,liang2025taming}. Related image anomaly detection methods use normalizing flows to model normal feature distributions on regular spatial grids~\cite{yu2021fastflow,zhou2024msflow}. More recently, flow matching has shown that transporting data to a simple Gaussian reference distribution provides a tractable density proxy, allowing low-density samples to be identified as anomalies~\cite{li2025and}. Despite these advances, a unified density-based approach for irregular point clouds remains lacking.

This gap stems from the assumptions underlying existing paradigms. Memory matching relies on a finite set of normal prototypes to cover potentially complex geometric variations. Consequently, uncovered yet valid patterns may lie far from all prototypes, while noisy local descriptors may yield unstable nearest-neighbor scores. Reconstruction assumes that a model trained only on normal data cannot faithfully recover anomalies. However, a high-capacity network may learn identity-like shortcuts and reconstruct abnormal regions. A distribution-based alternative is promising, but direct density modeling of unordered and sparsely sampled points is poorly suited to the regular feature grids required by efficient spatial flow models. Rendering an object from multiple views alleviates this representation mismatch but introduces another challenge: anomaly evidence produced in view-domain feature maps must be transferred to the corresponding original points while preserving visibility and projection correspondences. Therefore, a tractable learned transport framework is needed to connect structured semantic features with faithful 3D point-level evidence.

Our observation is that irregular 3D representation and structured density modeling need not operate in the same domain. Deterministic projection converts a point cloud into regular multi-view images, enabling mature 2D encoders to extract spatially organized semantic features and flow matching to model their normal distribution. Meanwhile, the projection establishes explicit geometric correspondences between each visible 3D point and its pixels across views. Using these correspondences as a bidirectional interface allows distribution evidence estimated in the 2D feature domain to be mapped back to point-level 3D scores, rather than treating rendered views merely as an auxiliary feature source.


Based on this observation, we introduce \textbf{Flow Matching for 3D Anomaly Detection (MVFM-3DAD)}, which reformulates unsupervised 3DAD as flow-guided density-proxy estimation in a multi-view semantic feature space. The Bidirectional Geometric Projector (BGP) renders a point cloud into structured views while retaining point-to-pixel correspondences. After anomaly evidence is estimated, the same correspondences guide its transfer and aggregation over the original 3D points. A frozen pretrained vision encoder extracts spatial semantic features, avoiding direct density estimation in raw coordinate space. The Flow-guided Density Proxy Estimator (FDPE) learns a time-conditioned, unidirectional transport from normal features to a standard Gaussian reference. During inference, test features follow the learned dynamics, and their terminal Gaussian energy measures compatibility with the transported normal distribution. This tractable score serves as a density proxy without requiring input reconstruction or explicit Jacobian or divergence evaluation. Aggregating local evidence across channels and views produces point-level anomaly scores, whose maximum yields the object-level prediction. We evaluate MVFM-3DAD on Real3D-AD~\cite{liu2023real3d} and MVTec3D-AD~\cite{bergmann2021mvtec} using object-level and point-level AUROC. MVFM-3DAD achieves the highest mean O-AUROC and P-AUROC on Real3D-AD, reaching $90.8\%$ and $96.0\%$, respectively. It also obtains the highest mean scores on MVTec3D-AD, with an O-AUROC of $95.9\%$ and a P-AUROC of $95.3\%$. These results demonstrate consistently strong performance across both benchmarks.


The main contributions of this work are summarized as follows:
\begin{itemize}
\item We formulate unsupervised 3DAD from a flow-guided density proxy perspective, assessing normality against a learned reference distribution rather than relying on finite memory matching or reconstruction-error scoring.
\item We introduce MVFM-3DAD, which establishes bidirectional projection between 3D point clouds and 2D images and estimates point-level density proxies from structured multi-view representations.
\item Extensive experiments demonstrate the effectiveness of MVFM-3DAD. It outperforms state-of-the-art methods by $8.5\%$ and $7.8\%$ in O-AUROC and P-AUROC on Real3D-AD, and by $0.6\%$ and $1.0\%$ on MVTec3D-AD.
\end{itemize}

\section{Related Works}\label{rw}
\subsection{3D Anomaly Detection}

3D anomaly detection aims to identify anomalous points or defective regions from point cloud data. Existing methods can be roughly divided into feature embedding methods and reconstruction-based methods\cite{liang2026mff,liang2025lightweight}. Feature embedding methods extract normal point cloud descriptors and detect anomalies by comparing test features with stored normal patterns. Representative methods include BTF \cite{horwitz2023back} and PatchCore \cite{roth2022towards}, which can be instantiated with different geometric descriptors or pretrained point features, and Reg3D-AD \cite{liu2023real3d}, which constructs memory banks from registered point cloud representations. M3DM \cite{wang2025m3dm} further extends this paradigm by introducing RGB-point cloud feature fusion, while CPMF \cite{cao2024complementary} and ISMP \cite{liang2025look} enhance 3D representations using complementary pseudo-modal information or spatial structural cues. Recent geometry-aware methods such as Curvature \cite{zha2026casl} and CASL \cite{zha2026casl} also exploit local surface properties to improve anomaly-sensitive feature learning.

Reconstruction-based methods learn to recover normal point cloud structures and regard regions with large reconstruction errors as anomalies. IMRNet \cite{li2024towards} reconstructs masked point regions, R3D-AD employs diffusion-based reconstruction, and DUS-Net \cite{liang2025taming} adopts a down-up sampling strategy to preserve group-level geometric centers. More recent methods such as SeDiR \cite{kim2026semantically} and AARD \cite{wu2026geometry} further improve reconstruction through semantic disentanglement or anomaly-aware diffusion guidance. Although these methods have achieved promising results, high-precision point clouds remain challenging due to large-scale point sets, complex geometry, and subtle local defects. 

\subsection{Flow-based Anomaly Detection}

Flow-based methods have been widely used in image anomaly detection to evaluate the likelihood or distributional compatibility. CFLOW-AD~\cite{gudovskiy2022cflow} first introduces conditional normalizing flows to image anomaly detection, while FastFlow~\cite{yu2021fastflow} adopts a fully convolutional flow architecture for efficient spatial density estimation. MSFlow~\cite{zhou2024msflow} further models multi-scale feature distributions with parallel flows and cross-scale fusion. However, normalizing flows require invertible transformations and tractable Jacobian determinants, which constrain architectural flexibility and distribution modeling.

Recent studies have explored flow matching and diffusion inversion as more flexible alternatives. WT-Flow~\cite{li2025and} transports normal data toward a Gaussian reference through time-reversed flow matching and introduces a worst-transport strategy to construct more discriminative trajectories in high-dimensional spaces. InvAD~\cite{sakai2026invad} replaces diffusion-based reconstruction with few-step DDIM inversion, directly mapping test features to the terminal noise distribution and measuring their latent typicality using spatial feature norms. Other approaches derive anomaly evidence directly from velocity-field behavior. TCCM~\cite{li2026scalable} detects tabular anomalies through their one-step deviation from a learned contraction trajectory. Flow Mismatching~\cite{chen2026flow} localizes anomalies by comparing the velocity predicted from normal data with the geometric velocity directed toward a test image. These methods demonstrate that continuous transport models support anomaly criteria beyond exact likelihood. However, they mainly operate on regular two-dimensional images or vectorized data.

\section{Methodology}\label{method}

\subsubsection{Problem Formulation}

Given a training set $\mathcal{D}_{tr}=\{\mathbf{P}_i\}_{i=1}^{M}$ containing only normal 3D point clouds, the goal of unsupervised 3DAD is to learn a parameterized detector $\Phi_{\theta}$ that can identify abnormal patterns in unseen test samples. For a test point cloud $\mathbf{P}\in\mathcal{D}_{te}$ with $N$ spatial points, the detector produces both an object-level anomaly score and a dense point-wise anomaly map:
$\Phi_{\theta}(\mathbf{P}) \rightarrow (s, \mathbf{S}),$
where $s\in[0,1]$ denotes the anomaly confidence of the entire object, and $\mathbf{S}=\{s_k\}_{k=1}^{N}$ represents the anomaly score for each 3D point. A larger $s_k$ indicates a stronger deviation of the $k$-th point from the learned normal geometric and appearance patterns.

\subsection{Overview}

\begin{figure}[!ht]
\centering
\includegraphics[width=1\columnwidth]{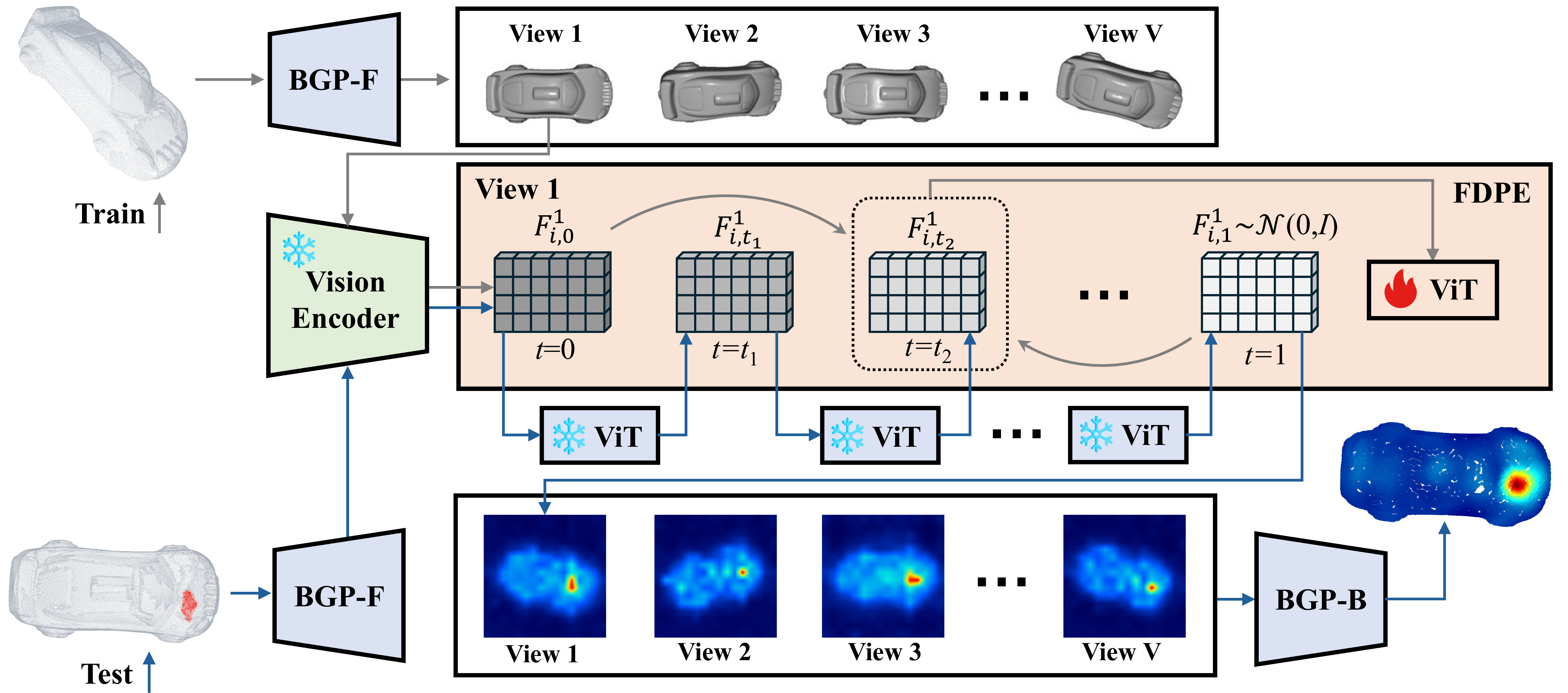} 
\caption{Overview of MVFM-3DAD. BGP projects 3D point clouds into multi-view images and maps anomaly evidence back to 3D points, while FDPE estimates anomaly scores through flow-guided density proxy modeling.}
\label{fig-ppline}
\end{figure}

In this work, we propose MVFM-3DAD, a novel framework for unsupervised 3DAD. As illustrated in Fig.~\ref{fig-ppline}, our method models the normality of 3D observations in a discriminative multi-view 2D feature space through flow matching. The overall framework consists of two core modules: Bidirectional Geometric Projector (BGP) and Flow-guided Density Proxy Estimator (FDPE).

\subsection{Bidirectional Geometric Projector}


Direct density estimation within the raw 3D input space presents a formidable challenge, as the observations inherently resides on complex low-dimensional manifolds and contains task-irrelevant variations. Image anomaly detection commonly addresses this issue with powerful pretrained encoders, whereas existing point descriptors mainly capture local geometry and provide limited global semantic context. Inspired by the 3D-to-2D projection strategy of CPMF~\cite{cao2024complementary}, the Bidirectional Geometric Projector (BGP) transforms an irregular point cloud into structured multi-view images, from which a pretrained Vision Transformer (ViT) extracts spatial semantic features. We denote this transformation as the forward process of BGP (BGP-F).

\noindent \textbf{Forward Projection:}
Given the $i$-th input point cloud
\begin{equation}
    \mathbf{P}_{i}
    =
    \left\{
        \mathbf{p}_{i,n}\in\mathbb{R}^{3}
    \right\}_{n=1}^{N_i},
\end{equation}
where $N_i$ is the number of valid points, we define a set of viewpoints
$\mathcal{V}=\left\{1,\ldots,V\right\}$.
Each viewpoint $v\in\mathcal{V}$ is associated with a rotation matrix
$\mathbf{R}^{v}\in\mathbb{R}^{3\times 3}$. In the forward process of BGP, each point is first rotated around the point-cloud center $\mathbf{c}_{i}$ to preserve the object location while changing its orientation:
\begin{equation}
    \widetilde{\mathbf{p}}_{i,n}^{v}
    =
    \mathbf{R}^{v}
    \left(
        \mathbf{p}_{i,n}-\mathbf{c}_{i}
    \right)
    +
    \mathbf{c}_{i},
    \qquad
    n=1,\ldots,N_i.
    \label{eq:bgp_rotation}
\end{equation}
After this view-specific rotation, BGP renders the point cloud into an image and records the projected pixel coordinate of every 3D point:
\begin{equation}
    \left(
        \mathbf{I}_{i}^{v},
        \mathbf{U}_{i}^{v}
    \right)
    =
    \operatorname{BGP}_{\mathrm{fwd}}^{v}
    \left(
        \mathbf{P}_{i};\mathbf{R}^{v}
    \right),
    \qquad
    \mathbf{U}_{i}^{v}
    =
    \left[
        \mathbf{u}_{i,1}^{v},\ldots,
        \mathbf{u}_{i,N_i}^{v}
    \right],
    \label{eq:bgp_forward}
\end{equation}
where $\operatorname{BGP}_{\mathrm{fwd}}^{v}(\cdot)$ denotes the forward projection under viewpoint $v$, including the rotation in Eq.~\eqref{eq:bgp_rotation} and the subsequent rendering.
Here, $\mathbf{I}_{i}^{v}\in\mathbb{R}^{3 \times H\times W}$ is the rendered image and $\mathbf{u}_{i,n}^{v}\in\mathbb{R}^{2}$ is the pixel coordinate associated with point $\mathbf{p}_{i,n}$.
A frozen pretrained ViT $f_{\phi}$ then encodes each view into a spatial feature map:
\begin{equation}
    \mathbf{F}_{i}^{v}
    =
    f_{\phi}
    \left(
        \mathbf{I}_{i}^{v}
    \right)
    \in\mathbb{R}^{d\times h\times w},
    \qquad
    v=1,\ldots,V.
    \label{eq:bgp_vit_feature}
\end{equation}
The resulting multi-view features provide semantic representations for density modeling, while the stored correspondences $\{\mathbf{U}_{i}^{v}\}_{v=1}^{V}$ are subsequently used to transfer the estimated anomaly evidence back to the original 3D points.

\noindent \textbf{Backward Projection:}
The forward projection preserves an explicit correspondence between each 3D point $\mathbf{p}_{i,n}$ and its projected coordinate $\mathbf{u}_{i,n}^{v}$ in every view. After FDPE produces the channel-wise anomaly evidence
$\mathbf{A}_{i}^{v}\in\mathbb{R}^{d\times h\times w}$, we first bilinearly upsample it to the rendering resolution, obtaining
$\widetilde{\mathbf{A}}_{i}^{v}\in\mathbb{R}^{d\times H\times W}$. BGP then samples the upsampled evidence at the stored projection coordinates and aggregates it across feature channels and viewpoints:
\begin{equation}
    s_{i,n}
    =
    \frac{1}{V}
    \sum_{v=1}^{V}
    \frac{1}{d}
    \sum_{\ell=1}^{d}
    \widetilde{A}_{i, \ell }^{v}
    \left[
        u_{i,n,y}^{v},
        u_{i,n,x}^{v}
    \right],
    \label{eq:bgp_backward}
\end{equation}
where $s_{i,n}$ denotes the anomaly score of point $\mathbf{p}_{i,n}$, and
$\mathbf{u}_{i,n}^{v}
=(u_{i,n,x}^{v},u_{i,n,y}^{v})^{\top}$.
Applying this operation to all valid points yields the point-level anomaly map
$\mathbf{s}_{i}=[s_{i,1},\ldots,s_{i,N_i}]^{\top}$.
This deterministic projection transfers multi-view anomaly evidence back to 3D without an additional learnable decoder.

\subsection{Flow-guided Density Proxy Estimator}
\label{sec:fdpe}

\begin{figure}[!ht]
\centering
\includegraphics[width=1\columnwidth]{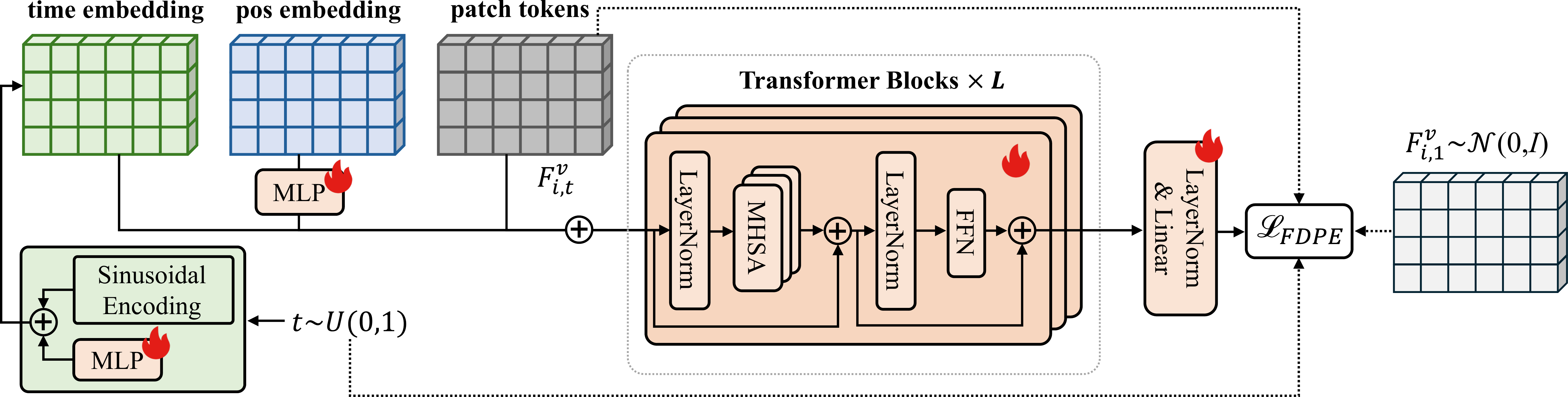} 
\caption{Overview of the proposed FDPE module. Patch tokens are derived from a frozen pretrained feature extractor.}
\label{fig-fdpe}
\end{figure}

Exact likelihood estimation with continuous flow models requires evaluating the divergence of the velocity field, which is computationally expensive for high-dimensional features. Following the density proxy perspective of WT-Flow~\cite{li2025and}, we instead learn to transport normal features toward a Gaussian reference distribution and measure their terminal Gaussian energy, as shown in Fig~\ref{fig-fdpe}. This formulation avoids explicit Jacobian estimation, thereby improving computational efficiency.

\noindent \textbf{Normal-feature transport.}
For each rendered view, BGP-F provides a semantic feature map
$\mathbf{F}_{i}^{v}\in\mathbb{R}^{d\times h\times w}$.
We first normalize the feature map
\begin{equation}
    \mathbf{X}_{i}^{v}
    =
    \mathrm{LN}
    \left(
        \mathbf{F}_{i}^{v}
    \right),
    \label{eq:fdpe_feature_norm}
\end{equation}
where $\mathrm{LN}(\cdot)$ denotes non-learnable layer normalization. We then sample a Gaussian endpoint
$\boldsymbol{\epsilon}\sim\mathcal{N}(\mathbf{0},\mathbf{I})$
with the same shape as $\mathbf{X}_{i}^{v}$, and construct a linear probability path from the normal feature to the Gaussian reference:
\begin{equation}
    \mathbf{X}_{i,t}^{v}
    =
    (1-t)\mathbf{X}_{i}^{v}
    +
    t\boldsymbol{\epsilon},
    \qquad
    t\in[0,1].
    \label{eq:fdpe_path}
\end{equation}
Along this path, the target transport direction is
\begin{equation}
    \boldsymbol{\delta}_{i}^{v}
    =
    \frac{d\mathbf{X}_{i,t}^{v}}{dt}
    =
    \boldsymbol{\epsilon}
    -
    \mathbf{X}_{i,0}^{v}.
    \label{eq:fdpe_target_velocity}
\end{equation}
FDPE parameterizes a time-conditioned velocity field
$g_{\psi}(\mathbf{X}_{i,t}^{v},t)$ and trains it only on normal samples by minimizing
\begin{equation}
    \mathcal{L}_{\mathrm{FDPE}}
    =
    \mathbb{E}_{\mathbf{X}_{i}^{v},\boldsymbol{\epsilon},t}
    \left[
        \left\|
            g_{\psi}
            \left(
                \mathbf{X}_{i,t}^{v},t
            \right)
            -
            \boldsymbol{\delta}_{i}^{v}
        \right\|_{2}^{2}
    \right].
    \label{eq:fdpe_loss}
\end{equation}
Since the training set contains only normal point clouds, the learned field captures how normal multi-view features move toward the Gaussian reference.

\noindent \textbf{Density-proxy inference.}
Given a test view feature $\mathbf{F}_{i}^{v}$, we initialize the ODE state by the normalized feature:
\begin{equation}
    \mathbf{Z}_{i,0}^{v}
    =
    \mathrm{LN}
    \left(
        \mathbf{F}_{i}^{v}
    \right).
    \label{eq:fdpe_initial_state}
\end{equation}
The learned transport is then solved from $t=0$ to $t=1$:
\begin{equation}
    \frac{d\mathbf{Z}_{i,t}^{v}}{dt}
    =
    g_{\psi}
    \left(
        \mathbf{Z}_{i,t}^{v},t
    \right).
    \label{eq:fdpe_ode}
\end{equation}
In practice, we use Euler integration with $N_{\mathrm{ode}}$ steps:
\begin{equation}
    \mathbf{Z}_{i,k+1}^{v}
    =
    \mathbf{Z}_{i,k}^{v}
    +
    \Delta t\,
    g_{\psi}
    \left(
        \mathbf{Z}_{i,k}^{v},t_{k}
    \right),
    \qquad
    \Delta t=\frac{1}{N_{\mathrm{ode}}},
    \quad
    t_{k}=k\Delta t.
    \label{eq:fdpe_euler}
\end{equation}
The terminal state $\mathbf{Z}_{i,N_{\mathrm{ode}}}^{v}$ is expected to be compatible with the Gaussian reference if the input feature follows the learned normal distribution. We therefore compute the channel-wise anomaly evidence as
\begin{equation}
    \mathbf{A}_{i}^{v}
    =
    1-
    \exp
    \left[
        -\,\frac{
            \left(
                \mathbf{Z}_{i,N_{\mathrm{ode}}}^{v}
            \right)^{2}
        }{2}
    \right]
    \in
    \mathbb{R}^{d\times h\times w}.
    \label{eq:fdpe_energy}
\end{equation}
A larger terminal energy indicates that the transported feature lies in a less typical region of the Gaussian reference distribution, and is therefore treated as stronger anomaly evidence. The resulting $\mathbf{A}_{i}^{v}$ is then passed to the backward process of BGP for point-level aggregation.

\noindent \textbf{Object-level score.}
After BGP-B transfers the view-domain evidence to the original 3D points, we obtain the point-level anomaly map
$\mathbf{s}_{i}=[s_{i,1},\ldots,s_{i,N_i}]^{\top}$.
The object-level anomaly score is computed by the maximum point response:
\begin{equation}
    s_{i}^{\mathrm{obj}}
    =
    \max_{1\leq n\leq N_i}
    s_{i,n}.
    \label{eq:fdpe_object_score}
\end{equation}

\section{Experiments}\label{exp}
\subsection{Experiment Settings}

\noindent \textbf{Datasets.} We evaluate our method on the Real3D-AD and MVTec3D-AD benchmarks. \textbf{Real3D-AD}~\cite{liu2023real3d} is a challenging point cloud anomaly detection benchmark collected from real industrial objects. It consists of 1,254 high-resolution 3D samples across 12 categories. Compared with RGB-D based 3D anomaly datasets, Real3D-AD provides denser point clouds and more complete 360-degree geometric information. \textbf{MVTec3D-AD}~\cite{bergmann2021mvtec} is a comprehensive benchmark for unsupervised 3D anomaly detection and localization. It contains 4,147 scans from 10 object categories. Each sample is represented by a three-channel coordinate image encoding the 3D surface geometry, and precise ground-truth annotations are provided for anomalous test samples.

\noindent \textbf{Baselines.} We compare our method with a comprehensive set of existing approaches, including IMRNet~\cite{li2024towards}, MVR-PCLIP~\cite{cheng2025toward}, PatchCore$_{\mathrm{ff}}$~\cite{roth2022towards}, BTF$_{\mathrm{raw}}$~\cite{horwitz2023back}, BTF$_{\mathrm{ff}}$~\cite{horwitz2023back}, M3DM~\cite{wang2025m3dm}, Reg3D-AD~\cite{liu2023real3d}, R3D-AD~\cite{zhou2024r3d}, CPMF~\cite{cao2024complementary}, ISMP~\cite{liang2025look}, DUS-Net~\cite{liang2025taming}, Curvature~\cite{zha2026casl}, CASL~\cite{zha2026casl}, SeDiR~\cite{kim2026semantically}, AARD~\cite{wu2026geometry}, MMRD~\cite{gu2024rethinking}, MVR-PCLIP~\cite{cheng2025toward}, EasyNet~\cite{chen2023easynet}, AST~\cite{rudolph2023asymmetric}, 3D-ADNAS~\cite{long2025revisiting}, Shape-Guided~\cite{chu2023shape}, 3DSR~\cite{zavrtanik2024cheating}, AnomalyCLIP~\cite{zhou2024anomalyclip}, PointAD~\cite{zhou2024pointad}, and HGCF~\cite{li2025hgcf}.

\noindent \textbf{Evaluation Metrics.}
We evaluate object-level anomaly detection using the area under the receiver operating characteristic curve (O-AUROC). For point-level anomaly localization, we report P-AUROC, computed over valid 3D points. 



\begin{table*}[!t]
\centering
\caption{Anomaly detection and localization results on the Real3D-AD benchmark, reported as O-AUROC/P-AUROC ($\uparrow$). Gray-shaded rows highlight our results, with \textbf{bold} values denoting the best results. ``--'' denotes unavailable results.}
\resizebox{\textwidth}{!}{
\begin{tabular}{l|c|ccccccccccccc}
\toprule
Method ($\downarrow$) & Venue ($\downarrow$) & Airpl. & Car & Candy & Chick. & Diamo. & Duck & Fish & Gemst. & Seaho. & Shell & Starf. & Toffe. & Mean \\
\midrule

IMRNet  & CVPR 24' &
76.2/-- & 71.1/-- & 75.5/-- & 78.0/-- & 90.5/-- & 51.7/-- & 88.0/-- & 67.4/-- & 60.4/-- & 66.5/-- & 67.4/-- & 77.4/-- & 72.5/-- \\
MVR-PCLIP  & TSMC 25' &
--/81.7 & --/69.5 & --/79.8 & --/88.8 & --/95.3 & --/88.1 & --/85.8 & --/91.5 & --/80.5 & --/90.1 & --/70.9  &--/93.4   &55.3/84.6   \\
\midrule
PatchCore$_{\mathrm{ff}}$  & CVPR'22 &
88.2/56.2 & 59.0/75.4 & 54.1/78.0 & 83.7/42.9 & 57.4/82.8 & 54.6/26.4 & 67.5/82.9 & 37.0/91.0 & 50.5/73.9 & 58.9/73.9 & 44.1/60.6 & 56.5/74.7 & 59.3/68.2 \\
BTF$_{\mathrm{raw}}$  & CVPR'23 &
73.0/56.4 & 64.7/64.7 & 53.9/73.5 & 78.9/60.9 & 70.7/56.3 & 69.1/60.1 & 60.2/51.4 & 68.6/59.7 & 59.6/52.0 & 39.6/48.9 & 53.0/39.2 & 70.3/62.3 & 63.5/57.1 \\
BTF$_{\mathrm{ff}}$  & CVPR'23 &
52.0/73.8 & 56.0/70.8 & 63.0/86.4 & 43.2/73.5 & 54.5/88.2 & 78.4/87.5 & 54.9/70.9 & 64.8/89.1 & 77.9/51.2 & 75.4/57.1 & 57.5/50.1 & 46.2/81.5 & 60.3/73.3 \\
M3DM  & CVPR'23 &
43.4/54.7 & 54.1/60.2 & 55.2/67.9 & 68.3/67.8 & 60.2/60.8 & 43.3/66.7 & 54.0/60.6 & 64.4/67.4 & 49.5/56.0 & 69.4/73.8 & 55.1/53.2 & 45.0/68.2 & 55.2/63.1 \\
Reg3D-AD  & NeurIPS'23 &
71.6/63.1 & 69.7/71.8 & 68.5/72.4 & 85.2/67.6 & 90.0/83.5 & 58.4/50.3 & 91.5/82.6 & 41.7/54.5 & 76.2/81.7 & 58.3/81.1 & 50.6/61.7 & 82.7/75.9 & 70.4/70.5 \\
R3D-AD  & ECCV'24 &
77.2/59.4 & 69.6/55.7 & 71.3/59.3 & 71.4/62.0 & 68.5/55.5 & 90.9/63.5 & 69.2/57.3 & 66.5/66.8 & 72.0/56.2 & 84.0/57.8 & 70.1/60.8 & 70.3/56.8 & 73.4/59.2 \\
CPMF  & PR' 24 &
70.1/61.8 & 55.1/73.4 & 55.2/83.6 & 50.4/55.9 & 52.3/75.3 & 58.2/71.9 & 55.8/98.8 & 58.9/44.9 & 72.9/96.2 & 65.3/72.5 & 70.0/80.0 & 39.0/95.9 & 58.6/75.8 \\
ISMP  & AAAI'25 &
85.8/75.3 & 73.1/83.6 & 85.2/90.7 & 71.4/79.8 & 94.8/92.6 & 71.2/87.6 & 94.5/88.6 & 46.8/85.7 & 72.9/81.3 & 62.3/83.9 & 66.0/64.1 & 84.2/89.5 & 75.7/83.6 \\
DUS-Net  & ACMMM'25 &
71.8/72.1 & 73.8/89.6 & 85.6/88.4 & 69.6/83.8 & 82.4/93.8 & 84.4/79.3 & 90.8/91.0 & 73.3/84.8 & 81.4/80.1 & \textbf{82.2}/87.2 & 75.5/79.9 & 83.4/86.1 & 79.5/84.7 \\
Curvature  & AAAI'26 &
34.5/71.5 & 68.6/69.9 & 88.9/81.0 & 49.6/70.5 & 95.7/86.0 & 76.8/86.0 & 85.4/72.7 & 55.4/90.4 & 79.0/54.5 & 59.4/61.4 & 87.2/51.9 & 87.2/79.7 & 72.3/72.9 \\
CASL  & AAAI'26 &
80.8/\textbf{84.2} & 79.9/90.5 & 84.8/93.2 & 65.7/71.3 & 97.6/98.8 & 83.6/89.5 & 93.5/93.5 & 76.9/91.6 & 64.3/81.4 & 79.1/87.3 & 89.3/83.9 & 92.4/93.7 & 82.3/88.2 \\
SeDiR  & CVPR'26 &
86.0/66.6 & 78.3/90.8 & 81.9/95.1 & 72.9/63.2 & 94.8/97.7 & \textbf{86.2}/83.4 & 93.8/96.7 & 62.7/54.7 & 67.4/73.9 & 77.9/79.8 & 85.4/70.9 & 84.5/94.0 & 81.0/80.6 \\
AARD  & CVPR'26 &
\textbf{88.2}/83.5 & 72.5/80.1 & 86.4/85.7 & 87.1/\textbf{90.6} & 91.7/88.6 & 84.6/89.4 & 89.1/91.7 & 76.5/92.6 & 80.7/85.2 & 81.1/84.6 & 75.9/78.5 & 85.5/80.6 & 82.0/86.0 \\

\rowcolor{gray!20} MVFM-3DAD & Ours &
67.6/83.7 &
\textbf{99.7}/\textbf{99.3} &
\textbf{89.9}/\textbf{97.2} &
\textbf{87.2}/89.4 &
\textbf{100.0}/\textbf{99.4} &
81.9/\textbf{97.5} &
\textbf{100.0}/\textbf{99.5} &
\textbf{94.1}/\textbf{98.4 } &
\textbf{97.4}/\textbf{96.9} &
80.6/\textbf{93.9} &
\textbf{92.0}/\textbf{97.6} &
\textbf{98.9}/\textbf{99.0} &
\textbf{90.8}/\textbf{96.0} \\
\bottomrule
\end{tabular}
}
\end{table*}

\noindent \textbf{Implementation Details.}
We train a model for each class using only normal training samples. All rendered views, organized point clouds, and ground-truth masks are resized to $224\times224$. Each object is represented by $27$ rendered views. We employ a frozen DINOv2 ViT-B/14 encoder as the feature extractor. The resulting tensor is normalized over its channel and spatial dimensions before being passed to FDPE. The velocity field is parameterized by an eight-block Vision Transformer with a hidden dimension of $768$, $12$ attention heads and global self-attention. The model is trained for $100$ epochs using AdamW with a learning rate of $1\times10^{-4}$ and weight decay of $1\times10^{-4}$. The learning rate is multiplied by $0.5$ every $25$ epochs. The DINOv2 encoder remains frozen, and only the velocity network is optimized. The batch size is set to 1. The learned flow is integrated with 20 Euler step. The resulting channel-wise energy maps are bilinearly upsampled to $224\times224$ and back-projected to valid 3D points. All experiments use a fixed random seed of $1$. The examined implementation evaluates the model every $10$ epochs. All experiments are implemented in PyTorch 2.6.0 and conducted on a single NVIDIA GeForce RTX 3090 GPU.

\subsection{Main Results on Real3D-AD}
MVFM-3DAD demonstrates a clear and consistent advantage in both anomaly detection and localization, achieving mean O-AUROC and P-AUROC scores of 90.8\% and 96.0\%, respectively, and outperforming the strongest competing method, CASL, by 8.5\% and 7.8\%. These margins substantially exceed the differences among the strongest baselines, confirming that the improvement is significant rather than marginal. MVFM-3DAD ranks first in 9 of 12 categories for O-AUROC and 10 of 12 categories for P-AUROC, with particularly large O-AUROC gains on Car, Gemstone, and Seahorse (19.8\%, 17.2\%, and 16.0\%) and P-AUROC gains on Starfish and Seahorse (13.7\% and 11.7\%). 
Collectively, these results show that the superiority of MVFM-3DAD is consistent across diverse categories rather than driven by a few exceptional cases.

\begin{table*}[!t]
\centering
\caption{Anomaly detection and localization results of the proposed method and other state-of-the-art methods on the MVTec3D-AD benchmark, reported as O-AUROC/P-AUROC ($\uparrow$). ``--'' denotes unavailable results.}
\label{tab:auroc_3d}
\resizebox{\textwidth}{!}{
\begin{tabular}{l|c|ccccccccccc}
\toprule
Method ($\downarrow$) & Venue ($\downarrow$) & Bagel & Gland & Carrot & Cookie & Dowel & Foam & Peach & Potato & Rope & Tire & Mean \\
\midrule
MMRD  & AAAI'24
& --/92.6 & --/80.6 & --/96.5 & --/85.8 & --/90.4 & --/73.1 & --/96.2 & --/95.8 & --/96.6 & --/93.6 & --/90.1 \\
LSFA  & ECCV'24
& --/97.4 & --/88.7 & --/98.1 & --/92.1 & --/90.1 & --/77.3 & --/98.2 & --/98.3 & --/95.9 & --/\textbf{98.1} & --/93.4 \\
MVR-PCLIP  & TSMC'25 &
--/97.1 & --/82.2 & --/94.4 & --/84.7 & --/72.0 & --/70.2 & --/98.5 & --/98.2 & --/96.6 & --/81.5 & 71.9/87.5  \\
EasyNet  & ACMMM'23
& 62.9/-- & 71.6/-- & 76.8/-- & 73.1/-- & 66.0/-- & 71.0/-- & 71.2/-- & 71.1/-- & 68.8/-- & 73.1/-- & 70.6/-- \\
AST  & WACV'23
& 88.1/-- & 57.6/-- & 96.5/-- & 95.7/-- & 67.9/-- & 79.7/-- & 99.0/-- & 91.5/-- & 95.6/-- & 61.1/-- & 83.3/-- \\
3D-ADNAS  & AAAI'25
& 79.4/-- & 65.6/-- & 85.9/-- & 79.5/-- & 78.0/-- & 62.9/-- & 84.3/-- & 78.1/-- & 82.4/-- & 87.8/-- & 78.4/-- \\

\midrule
BTF$_{\mathrm{ff}}$  & CVPR'23
& 82.5/97.3 & 55.1/87.9 & 95.2/98.2 & 79.7/90.6 & 88.3/89.2 & 58.2/73.5 & 75.8/97.7 & 88.9/98.2 & 92.9/95.6 & 65.3/96.1 & 78.2/92.4 \\
M3DM  & CVPR'23
& 94.1/94.3 & 65.1/81.8 & 96.5/97.7 & 96.9/88.2 & 90.5/88.1 & 76.0/74.3 & 88.0/95.8 & 97.4/97.4 & 92.6/95.0 & 76.5/92.9 & 87.4/90.6 \\
Shape-Guided  & ICML'23
& 98.3/97.4 & 68.2/87.1 & 97.8/98.1 & \textbf{99.8}/92.4 & 96.0/89.8 & 73.7/77.3 & 99.3/97.8 & 97.9/98.3 & 96.6/95.5 & 87.1/96.9 & 91.5/93.1 \\
CPMF  & PR'24
& 98.3/95.8 & 88.9/94.6 & 98.9/97.9 & 99.1/86.8 & 95.8/89.7 & 80.9/74.6 & 98.8/98.0 & 95.9/98.1 & 97.9/96.1 & \textbf{96.9}/97.7 & 95.1/92.9 \\
3DSR  & WACV'24
& 94.5/92.2 & 83.5/87.2 & 96.9/98.4 & 85.7/85.9 & 95.5/94.0 & 88.0/71.4 & 96.3/97.0 & 93.4/97.8 & \textbf{99.8}/\textbf{97.7} & 88.8/95.8 & 92.2/90.7 \\
AnomalyCLIP  & ICLR'24
& 81.2/89.1 & 65.7/94.6 & 70.5/98.0 & 68.5/82.5 & 58.7/93.6, & 67.8/\textbf{89.6} & 76.2/92.5 & 66.7/97.8 & 38.8/94.1 & 59.2/93.3 & 65.3/92.5     \\
PointAD  & NeurIPS'24
& 90.1/95.0 & 64.9/\textbf{95.3} & 94.7/\textbf{99.1} & 67.4/87.6 & 73.0/94.3 & 70.2/86.1 & 91.0/96.4 & 90.4./\textbf{99.2} & 86.8./97.1 & 60.0/93.1  & 78.9/94.3 \\
HGCF  & ACMMM'25
& 98.6/96.7 & 89.2/88.5 & 99.1/98.5 & 95.1/92.3 & 85.6/88.5 & \textbf{93.8}/85.1 & 99.2/98.5 & 97.5/98.1 & 99.5/96.1 & 95.2/87.2 & 95.3/93.0 \\

\rowcolor{gray!20} MVFM-3DAD & Ours
& \textbf{99.5}/\textbf{99.4}
& \textbf{97.6}/91.7
& \textbf{99.6}/96.6
& 99.0/\textbf{99.0}
& \textbf{97.3}/\textbf{94.5}
& 83.4/86.4
& \textbf{99.3}/\textbf{99.4}
& \textbf{98.2}/98.4
& 97.6/96.6
& 87.7/90.6
& \textbf{95.9}/\textbf{95.3} \\
\bottomrule
\end{tabular}
}
\end{table*}

\begin{figure}[!t]
\centering
\includegraphics[width=\columnwidth]{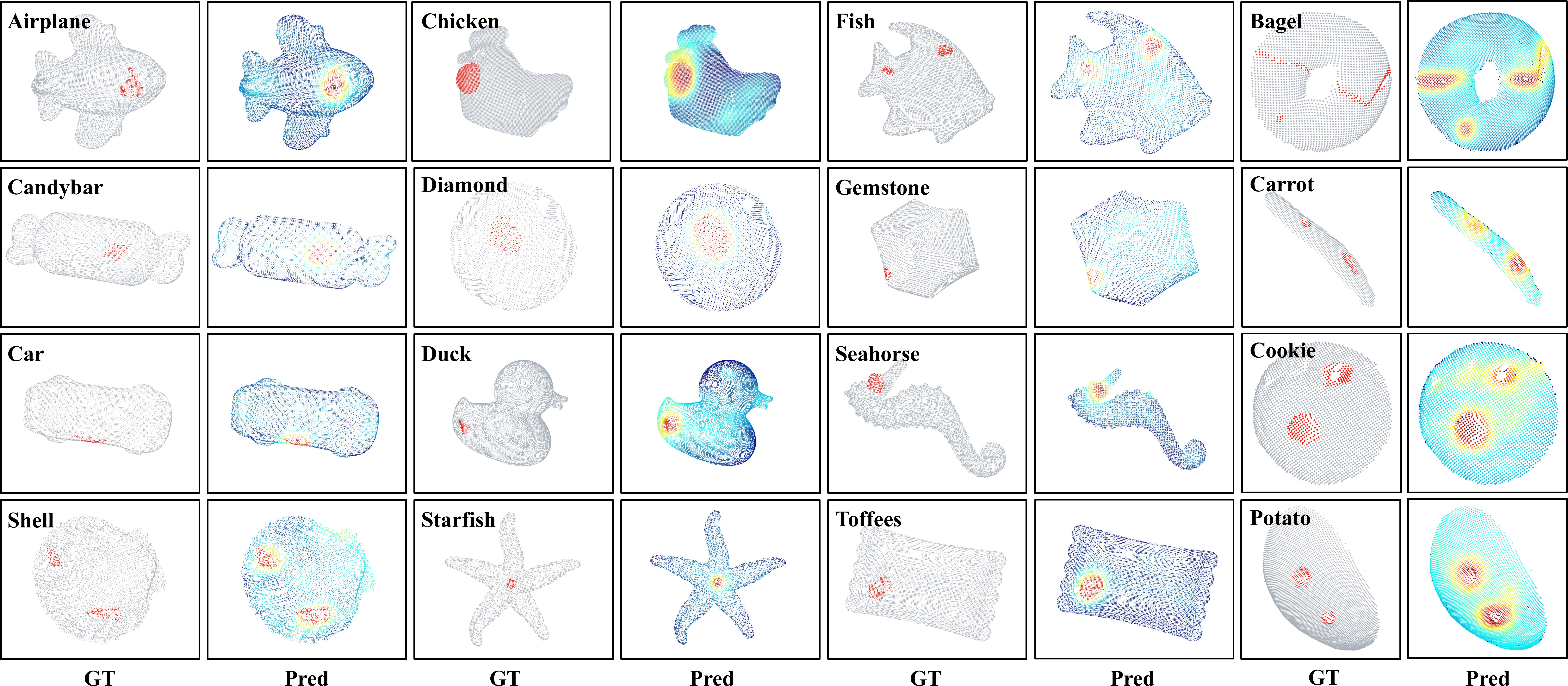} 
\caption{Qualitative results on the Real3D-AD and MVTec3D-AD benchmark.}
\label{fig-vis1}
\end{figure}

\subsection{Main Results on MVTec3D-AD}

MVFM-3DAD achieves the highest mean O-AUROC and P-AUROC of $95.9\%$ and $95.3\%$, outperforming the strongest competitors by $0.6\%$ and $1.0\%$, respectively. It ranks first in more than half of the evaluations for all the tasks, with notable detection gains on Dowel and Gland and localization gains on Foam and Bagel. MVFM-3DAD also reaches P-AUROC scores of $99.4\%$ on Peach and $99.0\%$ on Cookie. Although its detection performance remains weaker on Foam and Tire, the overall results demonstrate robust object-level detection and point-level localization across diverse geometries, supporting the effectiveness of the proposed density-proxy formulation.

\subsection{Qualitative Results}

To intuitively evaluate the anomaly localization performance of the proposed method, we conduct qualitative experiments on the Real3D-AD and MVTec3D-AD dataset, as illustrated in Fig.~\ref{fig-vis1}. The results show that our method produces clear and well-localized anomaly responses across diverse object categories, demonstrating its effectiveness in accurately identifying anomalous regions.

\subsection{Ablation Study}

As shown in Table~\ref{tab:ab}, we compare three detection paradigms using BGP and different feature representations under FDPE. FDPE improves over RD by $5.2\%$/$4.8\%$ and over MB by $27.5\%$/$19.0\%$ in O-AUROC/P-AUROC, demonstrating the advantage of density proxy estimation. With FDPE fixed, BGP outperforms FPFH by $9.7\%$/$10.6\%$ and PointMAE by $14.0\%$/$15.1\%$, indicating that structured multi-view features are better suited to density estimation than the evaluated 3D representations.

\begin{table*}[!t]
\centering
\caption{Ablation study on Real3D-AD in terms of O-AUROC and P-AUROC. MB, Recon., and DPE denote memory-bank-based, reconstruction-based, and density-proxy estimation paradigms. 2D and 3D indicate the corresponding input-space levels.}
\label{tab:ab}
\resizebox{0.65\textwidth}{!}{
\begin{tabular}{lccc|cc|cc}
\toprule
methods  &MB & Recon. & DPE &3D  &2D  & O-AUR  & P-AUR   \\
\midrule

BGP + PatchCore   &\checkmark   &   &  &  &\checkmark     &63.3   &77.0     \\
BGP + RD          &   &\checkmark   &  &  &\checkmark     &85.6   &91.2     \\
PointMAE + FDPE   &   &   &\checkmark  &\checkmark  &     &76.8   &80.9     \\
FPFH + FDPE       &   &   &\checkmark  &\checkmark  &     &81.1   &85.4     \\
BGP + FDPE        &   &   &\checkmark  &  &\checkmark     &90.8   &96.0     \\
\bottomrule
\end{tabular}
}
\end{table*}

\begin{figure}[!t]
\centering
\includegraphics[width=0.95\columnwidth]{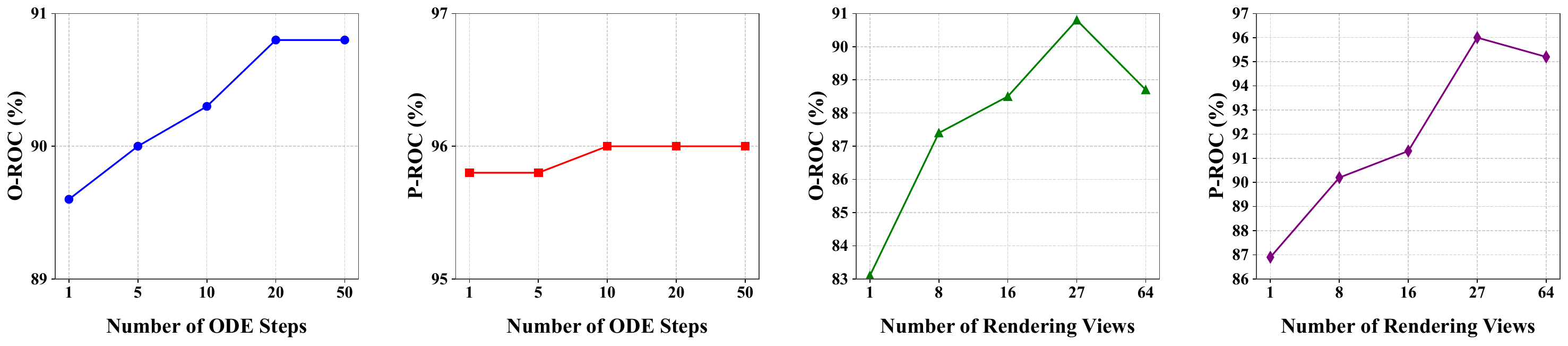} 
\caption{Hyperparameter analysis on the Real3D-AD benchmark.}
\label{fig-hpab}
\end{figure}

\subsection{Analysis of hyperparameters}
We quantitatively analyze the impact of two key parameters in our framework, the number of views rendered by the BGP and the inference ODE steps utilized in the FDPE. The results in Fig~\ref{fig-hpab} show that increasing the number of steps steadily improves detection and localization performance, which stabilizes after 20 steps. Increasing the number of views also yields substantial gains, with both metrics peaking at 27 views. Performance declines slightly at 64 views, suggesting that excessive views introduce redundant information.


\section{Conclusion}\label{conc}
This study shows that distribution consistency in a structured feature space provides a viable criterion for detecting and localizing geometric defects in point clouds. Results on Real3D-AD and MVTec3D-AD confirm the value of combining multi-view evidence with learned normal-feature transport, particularly for point-level localization. Future work will investigate more efficient view selection and improve robustness to large pose, sampling, and geometric variations.

\bibliographystyle{splncs04}
\bibliography{mybibliography}

\end{document}